\documentclass{iau} 
\usepackage{graphicx}

\title[Dissolution of star clusters harboring black hole subsystem] 
{MOCCA-SURVEY Database I: Dissolution \\ of tidally filling star clusters harboring \\ black hole subsystem}

\author[Mirek Giersz et al.]   
{Mirek Giersz$^1$, Abbas Askar$^2$, Long Wang$^{3,4,5}$, Arkadiusz Hypki$^6$, \\ Agostino Leveque$^1$ \and Rainer Spurzem$^{7,8,9}$ 
}

\affiliation{$^1$Nicolaus Copernicus Astronomical Centre, Polish Academy of Sciences, Warsaw, Poland \\ email: {\tt mig@camk.edu.pl} \\[\affilskip]
$^{2}$Lund Observatory, Department of Astronomy, and Theoretical Physics, Lund University, Box 43, SE-221 00 Lund, Sweden \\ email: {\tt askar@astro.lu.se} \\[\affilskip]
$^{3}$ Helmholtz-Institut f{\"u}r Strahlen- und Kernphysik, University of Bonn, Nussallee 14-16, D-53115 Bonn, Germany \\[\affilskip]
$^{4}$Argelander Institut F{\"u}r Astronomie, Auf Dem H{\"u}gel 71, 53121, Bonn, Germany\\
$^{5}$ RIKEN Center for Computational Science, 7-1-26 Minatojima-minami-machi, Chuo-ku, Kobe, Hyogo 650-0047, Japan \\ email: {\tt 
longwang.astro@live.com} \\[\affilskip]
$^{6}$Astronomical Observatory Institute, Faculty of Physics, A. Mickiewicz University, S\l{}oneczna 36, 60-286 Pozna\'n, Poland \\ email: {\tt ahypki@amu.edu.pl} \\[\affilskip]
$^{7}$National Astronomical Observatories and Key Laboratory for Computational Astrophysics, Chinese Academy of Sciences, 20A Datun Road, Chaoyang District, Beijing 100012, China\\
$^{8}$Astronomisches Rechen-Institut, Zentrum f{\"u}r Astronomie, University of Heidelberg, M¨onchhofstrasse 12-14, D-69120 Heidelberg, Germany\\
$^{9}$Kavli Institute for Astronomy and Astrophysics, Peking University, Yi He Yuan Lu 5, HaiDian District, Beijing 100871, China \\ email: {\tt spurzem@ari.uni-heidelberg.de}}

\pubyear{2019}
\volume{351}  
\jname{Star Clusters: From the Milky Way to the Early Universe}
\editors{A. Bragaglia, M.B. Davies, A. Sills \& E. Vesperini, eds.}
\begin{document}

\maketitle

\begin{abstract}
We investigate the dissolution process of star clusters embedded in an external tidal field and harboring a subsystem of stellar-mass black hole. For this purpose we analyzed the MOCCA models of real star clusters contained  in the Mocca Survey Database I. We showed that the presence of a stellar-mass black hole subsystem in tidally filling star cluster can lead to abrupt cluster dissolution connected with the loss of cluster dynamical equilibrium. Such cluster dissolution can be regarded as a third type of cluster dissolution mechanism. We additionally argue that such a mechanism should also work for tidally under-filling clusters with a top-heavy initial mass function.
\keywords{stellar dynamics, methods: numerical, globular clusters: general,   stars: black holes}
\end{abstract}

\firstsection 
\section{Introduction}

There are generally two types of cluster dissolution mechanisms discussed in the literature. One, slow, connected with relaxation-driven mass loss, and second, abrupt, connected with stellar evolution mass loss (e.g., \cite[Contenta \etal\ 2015]{Contenta2015}). \cite[Fukushige \& Heggie (1995)]{FukushigeHeggie1995} attributed the final abrupt cluster dissolution to the loss of dynamical equilibrium within the cluster. In such a situation, the cluster will not undergo core collapse. Many authors pointed out additional physical processes (apart from relaxation or stellar evolution),  which can influence the way in which a cluster is dissolved. In the 90s, \cite[Chernoff \& Weinberg (1990)]{ChernoffWeinberg1990} showed that star cluster structure, and the tidal field of a parent galaxy have a very strong influence on cluster evolution and its dissolution time. This was a seminal study, which later resulted in many more detailed papers discussing the final state of star clusters. It was shown that several other factors can also influence the lifetime of star clusters, such as: crossing time-scale (\cite[Whitehead \etal\ 2013]{Whitehead2013}), primordial binary population (\cite[Tanikawa \& Fukushige 2009]{TanikawaFukushige2009}), type of the Galactic orbit (\cite[Baumgardt \& Makino 2003]{BaumgardtMakino2003}), the form of the Galactic potential and tidal shocking (\cite[Gnedin \& Ostriker 1997]{GnedinOstriker1997}), and the properties of dark remnants retained in star clusters (\cite[Banerjee \& Kroupa 2011]{BanerjeeKroupa2011} and \cite[Contenta \etal\ 2015]{Contenta2015}). Inspired by those works, we decided to look at the cluster dissolution process in the models from the MOCCA Survey Database I (\cite[Askar \etal\ 2017]{Askar2017}), models of real star clusters simulated by the MOCCA code (\cite[Giersz \etal\ 2013]{Giersz2013} and \cite[Hypki \& Giersz 2013]{HG2013}). We introduced a third mechanism for cluster dissolution, which operates in tidally filling star clusters harboring a black hole subsystem (BHS). The physical processes responsible for the abrupt cluster dissolution are connected with strong energy generation in dynamical interactions between black holes (BH) and BH-BH binaries (BBHs) and strong cluster tidal stripping. The details of the third cluster dissolution mechanism are described by \cite[Giersz \etal\ (2019)]{Giersz2019}. Here, we will only provide short a summary and point out for the most important ingredients of the process of cluster dissolution. 

\begin{figure}[bht!]
 \vspace*{0.9 cm}
\begin{center}
 \includegraphics[width=0.65\textwidth,bb= 100 100 604 820,angle=-90]{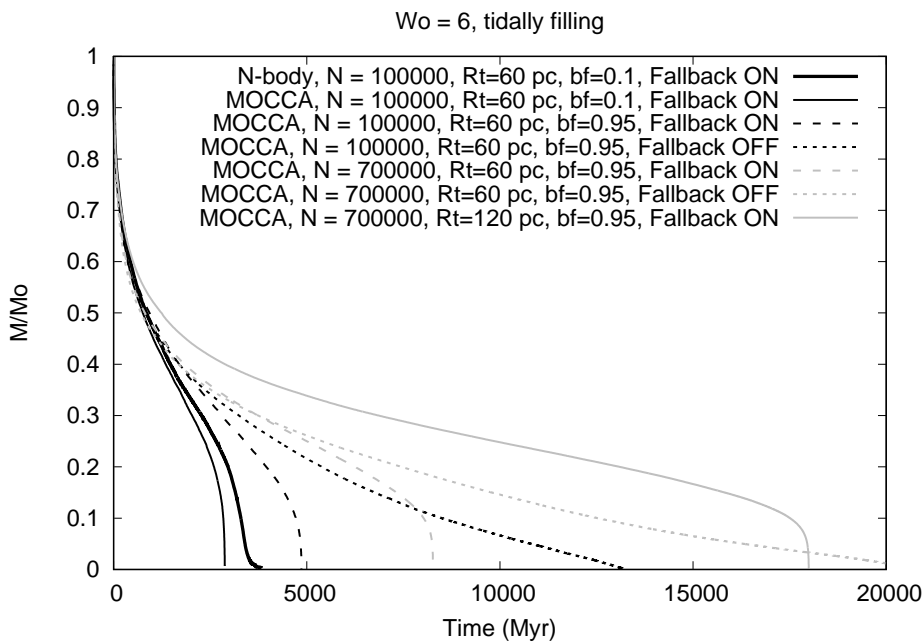} 
 \vspace*{-0.9 cm}
 \caption{Evolution of the fraction of cluster bound mass as a function of time for the MOCCA tidally filling models with King parameter, $W_0$, and different number of stars, $N$, tidal radii, $R_t$, mass FALLBACK ON or OFF. N-body model for $N=100000$ is depicted with thick black line. Description of models and lines are given inside the figure.}
   \label{fig1}
\end{center}
\end{figure}

\section{Results}

Cluster dissolution powered by stellar evolution mass loss coupled with strong cluster expansion and overflow of the tidal radius, $R_t$, is characterized by the abrupt loss of equilibrium and very fast dissolution. Usually, such clusters do not enter the core collapse phase of evolution. Dynamical evolution of models which are more concentrated is controlled by the relaxation process, not by stellar evolution mass loss. They evolve much more slowly and can enter the core collapse phase, provided that the relaxation was fast enough. The third cluster dissolution mechanism is a mixture of the above two processes - cluster enters the core collapse and post-collapse phases of evolution, and yet undergoes the abrupt final dissolution. The cluster core collapse is connected with strong mass segregation and BHS formation, and is followed, for usually a few Gyrs, by the balanced evolution between the BHS and the rest of the cluster (\cite[Breen \& Heggie 2013]{BreenHeggie2013}), and then abrupt dissolution. 

Fig. 1 shows  evolution of clusters which can form BHS (SNe kick velocities are reduced by mass fallback according to \cite[Belczynski \etal\ (2002)]{Belczynski2002} prescription) and clusters which do not form BHS because of high SNe kick velocities. Models have different numbers of objects, different tidal radii and binary fractions. Additionally, it shows an N-body model\footnote{N-body simulation was done by Long Wang with the use of the Nbody6++GPU code (\cite[Wang \etal\ 2015]{Wang2015}) with initial conditions exactly the same as for the MOCCA model} to check if MOCCA models properly account for the final stages of the cluster evolution. The N-body model shows a very similar shape of abrupt cluster dissolution as the MOCCA models, which clearly confirms that the abrupt dissolution of clusters observed in the MOCCA simulations are real features of cluster evolution. 

It is clear from Fig.1 that the abrupt dissolution does not depend on the global cluster properties, like tidal radius, cluster concentration, binary fraction or number of objects. They only determine the speed of the cluster evolution, as the relaxation time scale directly depends on those parameters. Generally, the larger the number of objects or the larger the binary fraction or the smaller the cluster concentration, the longer the abrupt dissolution time. It is worth pointing out that the observed behavior of cluster dissolution happens only for tidally filling models (the ratio between tidal and half-mass radii is typically about 7 for King model with $W_0=6$), models which do not have the room to freely expand. This, together with the fact that BHS is present until the final stages of cluster dissolution suggests that BHS and strong tidal stripping have a crucial role in destroying the balanced cluster evolution and lead to abrupt cluster dissolution connected with the loss of cluster dynamical equilibrium. 

As it is clearly shown by \cite[Giersz \etal\ (2019)]{Giersz2019} cluster dissolution is connected with the decoupling of the BHS from the rest of the cluster and the loss of dynamical equilibrium by other objects. Luminous objects become hot and expand. BHS stops to collapse further and slowly starts to expand and disrupts itself like an isolated cluster. This picture further supports the evolution of the escape velocity and the escape rate for tidally filing models. The cluster is constantly tidally stripping, but the internal cluster structure is not strongly changing, so the escape velocity becomes smaller and smaller, and more and more objects can be kicked out from the cluster. The positive feedback between strong energy generation, high flow of escaping stars, decreasing escape velocity and enhanced tidal stripping leads finally to the loss of cluster dynamical equilibrium and abrupt dissolution. 

The above described third mechanism of cluster dissolution is universal and should work for all clusters which harbor BHSs to the end of their life, and are exposed to a strong tidal field, regardless of the initial cluster structure or surrounding environment.

\section{Conclusions}

Tidally filling clusters with BHS in the center can show abrupt dissolution, provided that the initial King concentration parameter is moderate and BHS can survive until cluster death. The dissolution is controlled by a strong energy generation in dynamical interactions inside the BHS and strong tidal stripping, which leads finally to sudden loss of dynamical equilibrium and decoupling of BHS from the rest of the cluster.
Such mechanism should operate also for tidally under-filling clusters with top-heavy initial mass functions. 

Just before the abrupt dissolution the cluster will look like a 'dark cluster' as described by \cite[Banerjee \& Kroupa (2011)]{BanerjeeKroupa2011}, which is a different kind of 'dark cluster', which harbors an IMBH with mass comparable to the cluster total mass (\cite[Askar \etal\ 2017]{Askar2017}). 

The fast dissolution of massive, tidally filling clusters can have strong influence on the estimated rate of BH-BH mergers from GCs. If clusters are born tidally filling and close to the Galactic center, then we can expect a lot of free floating BHs in the Galactic Bulge.

\section*{Acknowledgements}

MG and AL were partially supported by the Polish National Science Center (NCN) through the grant UMO-2016/23/B/ST9/02732. AA is currently supported by the Carl Tryggers Foundation for Scientific Research through the grant CTS 17:113. LW in this work was supported by the funding from Alexander von Humboldt Foundation. AH was supported by Polish National Science Center grant 2016/20/S/ST9/00162. RS has been supported by the National Natural Science Foundation of China (NSFC), grant 11673032. We acknowledge support from the Chinese Academy of Sciences through the Silk Road Project at the National Astronomical Observatories, Chinese Academy of Sciences (NAOC), through the 'Qianren' special foreign experts programme of China, and through the Sino-German collaboration program GZ1284. This work benefited from support by the International Space Science Institute (ISSI), Bern, Switzerland, through its International Team programme ref. no. 393 The Evolution of Rich Stellar Populations \& BH Binaries (2017-18).

\end{document}